\documentclass[onecolumn]{article}              
\usepackage{authblk}
\usepackage{graphicx}
\usepackage[square,comma,numbers]{natbib}
\usepackage{a4wide}
\usepackage{float}

\usepackage{array}
\newcolumntype{L}[1]{>{\raggedright\let\newline\\\arraybackslash\hspace{0pt}}m{#1}}
\newcolumntype{C}[1]{>{\centering\let\newline\\\arraybackslash\hspace{0pt}}m{#1}}
\newcolumntype{R}[1]{>{\raggedleft\let\newline\\\arraybackslash\hspace{0pt}}m{#1}}

\begin{document}

\title{PyNX: high performance computing toolkit for coherent X-ray imaging based on operators}

\author[a,b]{Favre-Nicolin, Vincent}
\author[a]{Girard, Ga\'etan}
\author[a]{Leake, Steven}
\author[c]{Carnis, J\'er\^ome}
\author[a]{Chushkin, Yuriy}
\author[a]{Kieffer, J\'er\^ome}
\author[a]{Pal\'eo, Pierre}
\author[b,d]{Richard, Marie-Ingrid}

\affil[a]{ESRF, The European Synchrotron, 71 Av. des Martyrs, 38000 Grenoble, France}
\affil[b]{Univ. Grenoble Alpes, Grenoble, France}
\affil[c]{Deutsches Elektronen-Synchrotron (DESY), D-22607, Hamburg, Germany}
\affil[d]{CEA, IRIS-MEM, Nanostructures and Synchrotron Radiation Laboratory, F-38000 Grenoble, France}


\maketitle 

\begin{abstract}
The open-source PyNX toolkit \cite{favre-nicolin_fast_2011, mandula_pynx.ptycho-_2016} has been extended to provide tools for coherent X-ray imaging data analysis and simulation. All calculations can be executed on graphical processing units (GPU) to achieve high performance computing speeds. This can be used for Coherent Diffraction Imaging (CDI), Ptychography and wavefront propagation, in the far or near field regime. Moreover, all imaging operations (propagation, projections, algorithm cycles..) can be used in Python as simple mathematical operators, an approach which can be used to easily combine basic algorithms in a tailored chain. Calculations can also be distributed to multiple GPUs, e.g. for large Ptychography datasets. Command-line scripts are also available for on-line CDI and Ptychography analysis, either from raw beamline datasets or using the Coherent X-ray Imaging data format \cite{maia_coherent_2012}.
\end{abstract}


\section{Introduction}

Coherent X-ray Imaging techniques have been intensely developed during the last 20 years thanks to the wide availability of synchrotron sources with high brilliance. This covers a wide range of techniques, beginning with phase contrast imaging \cite{cloetens_phase_1996, nugent_quantitative_1996, cloetens_holotomography-_1999}, coherent diffraction imaging \cite{miao_extending_1999,miao_approach_2001, marchesini_x-ray_2003}, allowing to reconstruct single objects from their diffraction pattern alone, including strain imaging of crystalline nano-objects in the Bragg geometry \cite{williams_three-dimensional_2003, pfeifer_three-dimensional_2006,robinson_coherent_2009, favre-nicolin_analysis_2010}. More recently X-ray Ptychography \cite{thibault_high-resolution_2008,thibault_probe_2009,maiden_improved_2009}, which can be used both in the far field and near field regime \cite{stockmar_near-field_2013,stockmar_x-ray_2015}, was developed for imaging extended objects (larger than the incident beam), both in the small angle and in the Bragg geometry \cite{chamard_strain_2015,hruszkewycz_high-resolution_2016}; this technique can also be used in the Fourier regime by scanning the transmitted beam \cite{wakonig_x-ray_2019}.

These techniques all provide high-resolution two or three-dimensional imaging, down to 5 to 15 nanometer resolution depending on the experimental setup. The main experimental requirement is a coherent X-ray beam, which is readily available at synchrotron facilities. These experiments will see the main benefit of the current upgrades of synchrotron rings, which promise two orders of magnitude increase in the available coherent X-ray flux thanks to higher brilliance \cite{johansson_max_2016, raimondi_esrf-ebs:_2016,schroer_hard_2016}, and will enable faster dynamics and imaging experiments, as well as reach higher resolution \cite{favre-nicolin_dynamics_2017}. The availability of a higher coherent fraction at high energy ($>$20keV) will enable data collection for thicker samples and allow to mitigate radiation damage with lower absorption.

Data analysis often remains a bottleneck for these experiments, either from the complexity of the algorithms, or simply the computing requirements. A variety of software is readily available for Ptychography \cite{nashed_parallel_2014, enders_computational_2016,marchesini_sharp-_2016,nashed_distributed_2017, odstrcil_iterative_2018, dong_high-performance_2018, wakonig_ptychoshelves_2020}, and fewer for CDI \cite{maia_hawk-_2010,newton_bonsu-_2012}. Several limitations remain: (i) software packages are not always publicly distributed, (ii) high-performance computing - generally based on graphical processing units (GPU) is not always available or complicated to setup, and (iii) the software can be difficult to maintain or improve due to the complexity of algorithms or their GPU implementation.

In this article we will present the open-source coherent X-ray imaging modules of the PyNX toolkit. In the previous versions \cite{favre-nicolin_fast_2011, mandula_pynx.ptycho-_2016}, GPU-accelerated computing was only available for scattering calculations (which are unchanged), whereas this new version is a complete rewrite of the Ptychography module, and adds tools for CDI and Wavefront calculations, all GPU-accelerated. We will first present an outline of the toolkit organisation, followed by details of the operator-based approach which is used to simplify the development of custom algorithms, and finally the available command-line scripts.


\section{PyNX toolkit organisation}

PyNX is written primarily in Python, using the NumPy and SciPy libraries \cite{virtanen_scipy_2019} for basic data processing it is organised in several modules for the different tasks:
\begin{itemize}
    \item \texttt{cdi}: Coherent Diffraction Imaging. See section \ref{sec_cdi}.
    \item \texttt{operator}: classes and functions for the \textit{operator}-based approach, which is described in section \ref{sec_cdi}.
    \item \texttt{processing\_unit}: functions to automatically select and initialise the processing units, either using OpenCL or CUDA.
    \item \texttt{ptycho}: Ptychography. See section \ref{sec_ptycho}.
    \item \texttt{scattering}: legacy GPU-accelerated kinematical scattering calculations, as described in \cite{favre-nicolin_fast_2011}.
    \item \texttt{test}: automated test routines and scripts.
    \item \texttt{utils}: array handling and plotting routines.
    \item \texttt{wavefront}: coherent wavefront propagation, mostly for simulation purposes.
\end{itemize}

\begin{figure}
\label{fig-profile}
\includegraphics[width=\textwidth]{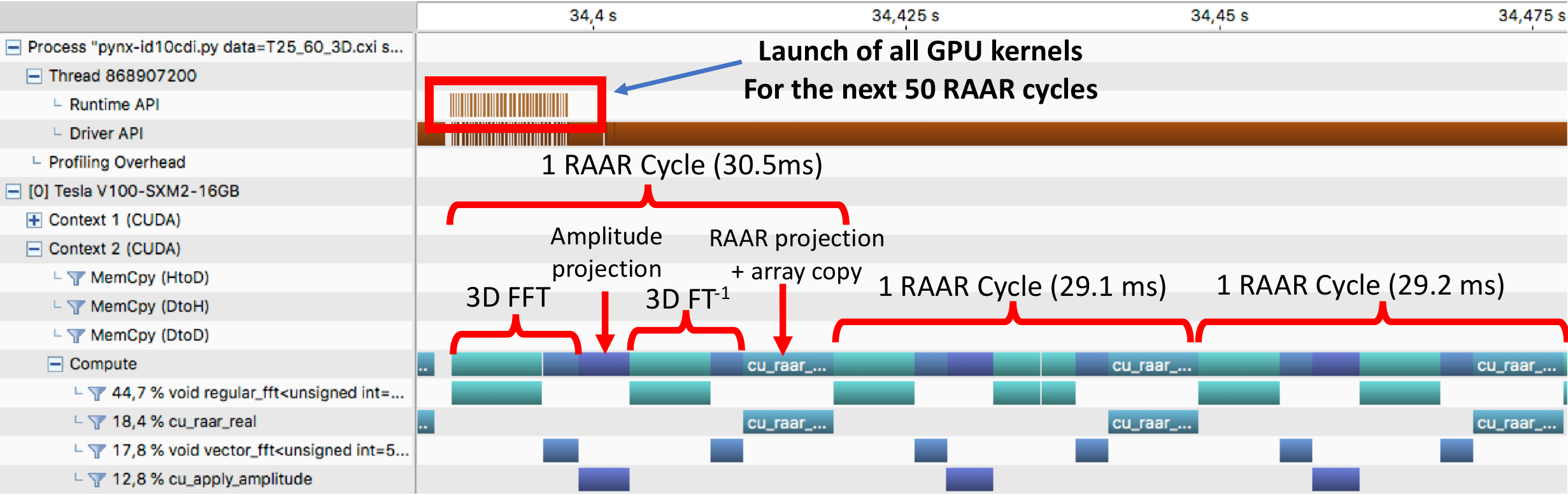}
\caption{Example profiling during a CDI analysis: the algorithm consists of 50 cycles of Relaxed Averaged Alternating Reflections (RAAR) \protect\cite{luke_relaxed_2005} applied to a $512^3$ voxels object. The process includes a Fourier Transform of the complex object $O(n)$, an amplitude projection to apply the observed amplitudes, an inverse Fourier transform, a linear combination with the previous object $O(n)$ depending on whether the object voxel is within the support. As can be seen from the nVidia visual profiler graphical display, the command for \protect\textit{all} the 50 cycles of RAAR are executed ('Runtime API') within 30 ms, and are then executed \protect\textit{asynchronously}: once the commands have been queued, the GPU is constantly working on the different parts of the algorithm (intervals in white or light grey indicate no activity), while the Python code can independently prepare other commands to be queued. This asynchronous behaviour is close to an optimal GPU performance. Such a calculation can be executed in Python simply by writing: \protect\texttt{ cdi = RAAR()**50 * cdi}.  }
\end{figure}

All calculations for the new coherent imaging modules (\texttt{cdi}, \texttt{ptycho} and \texttt{wavefront}) are executed on the available GPU, either using the pyCUDA or pyOpenCL libraries \cite{klockner_pycuda_2009}. The language is automatically selected as well as the GPU: this is done by favouring CUDA over OpenCL if both are available (see supplementary figures \ref{fig:suppl1} and \ref{fig:suppl2} in the appendix for a comparison), and then selecting the fastest GPU (based on a 2D FFT test) if more than one is available. The user can also opt to select a language and/or a GPU from its name or rank among those available.

One important aspect of the code design is that while most (84\% lines) of it is written in Python, all the algorithms are executed \textit{asynchronously} on the GPU, i.e. the commands are sent by the Python process to the GPU for a large number of operations\footnote{Note that GPU kernels are naturally executed asynchronously, but in order to maintain a high performance, it is important to design all algorithms to fetch values from the GPU as scarcely as possible (every 10s or 100s of cycles), so that the GPU process is not slowed down by a synchronisation process, even if it is only to copy a few floating-point values}. This ensures that the execution on the GPU is rarely interrupted, i.e. only when it is necessary to get back data from the GPU to the computer main memory. 

Illustrated in figure \ref{fig-profile}, the graphical view of the profiling of a CDI optimisation is demonstrated. It can be seen that after launching the chain of algorithms from Python, the code is executed on the GPU with negligible latency between the different operations (Fourier transforms, support projections, copy of arrays, ...). This remains true until some data has to be retrieved from the GPU memory, but only happens when fetching data from the GPU (e.g. for plotting or displaying figures of merit).

Finally, the \verb cdi  and \verb ptycho  modules also include a \verb runner  sub-module, which handles automated data processing using command-line scripts, which will be described in section \ref{sec_cli}.


\section{\label{sec_cdi}Coherent Diffraction Imaging and operators}

The CDI technique consists of reconstructing an object from a far-field diffraction pattern alone \cite{miao_extending_1999}, a technique which has been expanded to three-dimensional reconstruction by collecting multiple ($>$100) projections around a rotation axis, either in the small angle \cite{chapman_high-resolution_2006, chushkin_three-dimensional_2014} or in the Bragg geometries \cite{williams_three-dimensional_2003, robinson_coherent_2009} - the latter approach also yields information about strain in the reconstructed object.

\begin{figure}
\begin{verbatim}
import numpy as np
# This imports all necessary operators & classes
# The GPU and language (OpenCL/CUDA) are auto-selected
from pynx.cdi import *
# Load data, support and mask of bad pixels
iobs = np.load("iobs.npz")['iobs']
mask = np.load("mask.npz")['mask']
support = np.load("support.npz")['support']
# Create main CDI object
cdi = CDI(iobs, support=support, mask=mask)
# Initial scaling of object with respect to Iobs
cdi = ScaleObj(method='F') * cdi
# Do 40 cycles of HIO with a positivity constraint
cdi = HIO(positivity=True)** 40 * cdi
# Support update operator
sup = SupportUpdate(threshold_relative=0.17)
# Do 20*(40 cycles of HIO, 5 of ER, support update)
cdi = ShowCDI()* (sup*ER()** 5*HIO()** 40)**20 * cdi
\end{verbatim}
    \caption{Example CDI reconstruction code using operators. All the reconstruction operations (\texttt{HIO}, \texttt{ER}, \texttt{SupportUpdate}) are transparently (no explicit initialisation is required) and asynchronously executed on the GPU. Only when the \texttt{ShowCDI} operator is used, the resulting object and support are fetched from the GPU for display, which automatically waits for all operations queued on the GPU to be finished.}
    \label{fig:fig_cdi_code}
\end{figure}

In order to reconstruct the object from non-redundant diffraction data it is necessary to recover the lost phases of the measured amplitudes. A variety of algorithms are available, all of which rely on alternating between a real-space estimate of the object, where its extent (the so-called 'support' of the object) is evaluated, and diffraction (Fourier) space, where an amplitude constraint can be applied from the measured intensities. A unified view of these algorithms has been presented in \cite{marchesini_unified_2007}, with a demonstration that all operations applied to the object array can be described as mathematical operators. For example, the simplest algorithm -error reduction- can be written in the following way:
\begin{equation}
    \rho^{(n+1)}=P_S\mathcal{F}^{-1}P_m\mathcal{F}\rho^{(n)}
\end{equation}
where $\rho^{(n)}$ is the object's complex density array at iteration $(n)$, $\mathcal{F}$ is the Fourier transform, $P_m$ is a magnitude projection operator which replaces the modulus of the calculated scattering by the observed ones, and $P_S$ is the support projection operator, which multiplies the object array by 0 outside its support.

Since a complete algorithmic chain used to retrieve an object relies on a large number (at least a few hundred) of such mathematical operations, it is convenient to use object-oriented programming to enable writing the sequence of operations exactly as mathematical operations. This is achieved in the following way:
\begin{enumerate}
    \item a CDI class is defined, including as data the object array (either in real or Fourier space) and the support array (0 outside the support, 1 inside), with a few input/output functions.
    \item a family of CDI operators is created, each operator allowing to alter or analyse a CDI object by left-multiplication. These operators also take care of preparing the data and execution kernels in GPU space.
\end{enumerate}
For example if $cdi$ is a CDI object, applying an Error Reduction algorithm one simply writes:
\begin{verbatim}
    cdi = ER() * cdi
\end{verbatim}

The main property of operators is that they can be multiplied by another operator to be chained, allowing arbitrarily long operations, or raised to a given integer N to execute the operator N consecutive times. For example:
\begin{verbatim}
    cdi = ER()**20 * HIO()**100 * cdi
\end{verbatim}
will apply 100 cycles of Hybrid Input-Output followed by 20 cycles of Error Reduction.

This operator-based approach presents several advantages: (i) it is very easy to combine and alter the algorithmic chain which is used for the data analysis, allowing greater flexibility for a wide range of datasets (choices may vary depending on signal/noise quality, hard or soft condensed matter, low or high energy, etc..) and (ii) operators are independent pieces of code which can very easily be expanded or replaced as needed, avoiding the risk of turning the program into a cathedral-like construction which cannot evolve. The only limit to that approach is that the way the arrays (object, observed intensity, masks) are stored in the \texttt{CDI} object must be common to all algorithms.

A list of the most important CDI operators is given in appendix \ref{appendix_cdi_operators}, and an example reconstruction code is given in figure \ref{fig:fig_cdi_code}. Such an operator-based approach combined with the asynchronous execution can achieve close to optimal speed, as illustrated in Figure \ref{fig-profile}, even when combining algorithms. 

In order to analyse the performance in detail, we can compare the time for a single RAAR cycle for a dataset of $512^3$ voxels. As on a GPU, the speed is generally limited by the memory transfers (and not the time to perform floating-point operations), the relevant figure is the bandwidth of the process. One RAAR cycle shown in Figure \ref{fig-profile} takes $\approx$29.5 ms, and requires 9.625 read and 9 writes \footnote{1 read + 1 write for each of the dimensions of the forward and backward FFT, 1 read and 1 write to apply the amplitude constraints to the complex array in Fourier space, 0.5 read of the intensity array (32-bit float instead of 32-bit complex), and 2.125 read + 2 writes for the RAAR projections operation (the .125 comes from reading the 8-bit support array)} of the 3D complex array in 32-bit precision. This corresponds to a memory throughput of 677 GB/s. This can be compared to the theoretical throughput of 900 GB/s, and the observed actual throughput for a simple on-GPU copy of 790 GB/s, which shows that the asynchronous execution delivers a very good performance. 

\begin{figure}
\label{fig-result-script-ptycho-cdi}
\includegraphics[width=1.0\textwidth]{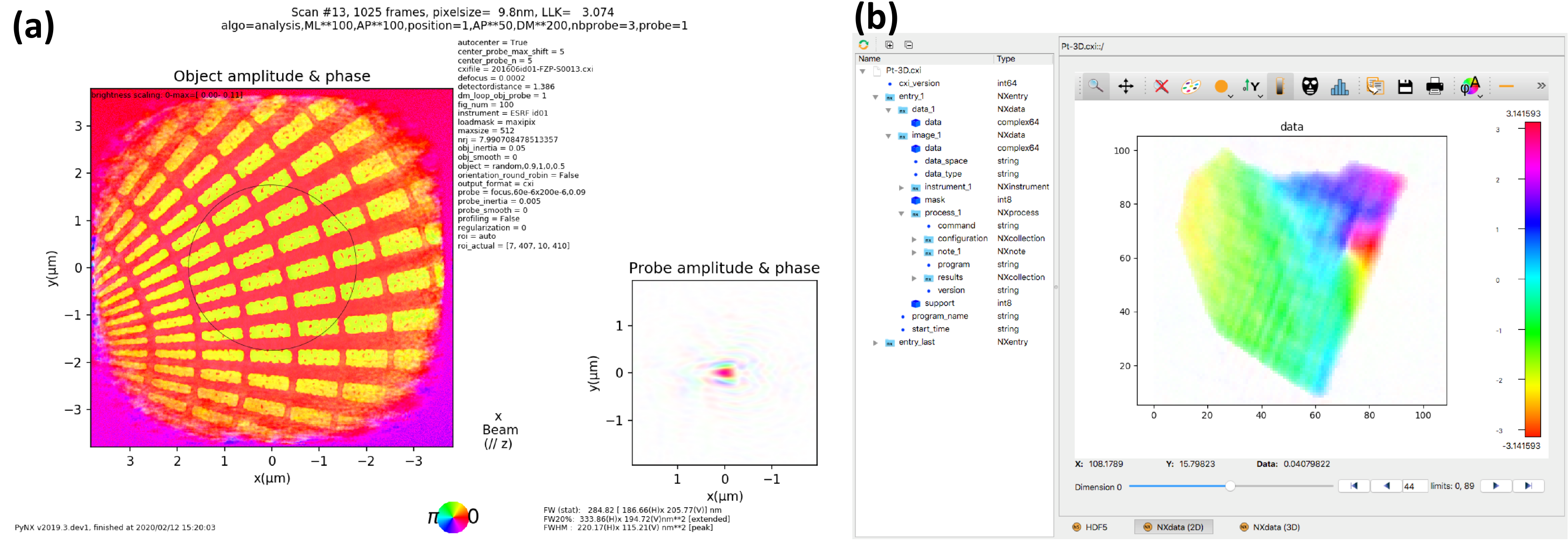}
\caption{Example results from a command-line analysis using PyNX: (a) plot from the Ptychography analysis of a Siemens star, with the object and the probe. This plot is automatically produced at the end of the script by the 'analysis' step, and includes the algorithm chain used for the analysis as well as all parameters, which are also saved in the CXI output file. (b) View of the result of a CDI analysis on a 3D Pt nano-particle with a dislocation, using the silx toolkit viewer \protect\cite{silx_silx-_2020, vincent_silx-_2020}. Note on the left the details of the file contents with CXI/HDF5/NeXuS formatting. In all graphics, a hue-saturation color scheme is used to represent both the amplitude as saturation, and the phase as color, as indicated by the colour wheel at the bottom of (a). Both figures are available in a larger version as supplementary figures \ref{fig:suppl3} and \ref{fig:suppl4} in the appendix}
\end{figure}

\begin{table}
\begin{scriptsize}
\begin{tabular}{|c|c|c|c|c|c|}      
\hline
 Method    & Configuration & Size of dataset        & Algorithm chain       & $<$time/cycle$> (ms)$ & $\Delta t_{total}$(s)    \\ \hline
 CDI       & \shortstack{support update \\ every 50 cycles}   & 512$\times$512$\times$512     & ER()**200 * RAAR()**600  &  27(ER) 30(RAAR)& 24    \\ \hline
 Ptychography & far field, 1 probe  & \shortstack{1000$\times$(256$\times$256)\\ object: 1176$\times$1188}      & ML()**100*DM()**100      & 17 (DM) 34 (ML) & 6.6    \\ \hline
 Ptychography & far field, 3 probe   & \shortstack{1000$\times$(256$\times$256)\\ object: 1176$\times$1188}      & ML()**100*DM()**100      & 44 (DM) 92 (ML) & 15    \\ \hline
 Ptychography & near field, 1 probe   & 17$\times$(2048$\times$2048)     & ML()**100*DM()**400      & 39 (DM) 84 (ML)  & 25   \\ \hline
 MPI-Ptychography & far field, 1 probe  & \shortstack{70.10$^3\times$(512$\times$512)\\ object: 16940$\times$16300}      & ML()**200*DM()**400      & 480 (DM) 890 (ML) & 409    \\ \hline
 MPI-Ptychography & far field, 1 probe  & \shortstack{250.10$^3\times$(256$\times$256)\\ object: 15653$\times$15179}   & ML()**200*DM()**400      & 375 (DM) 760 (ML) & 329    \\ \hline
\end{tabular}
\end{scriptsize}

\caption{\label{table_speed}Example speed achieved using a single nVidia V100 GPU (or 12 GPU using MPI for the last two lines), for coherent diffraction imaging and Ptychography data analysis, using the CUDA language. The total time reported only include algorithm time, not loading or saving results, but includes some overhead (typically 5-20\%) compared to individual cycles, due to fetching data from the GPU (log-likelihood reporting), initialising GPU kernels, or extra operations (scaling, check for drifts,..). The CDI algorithm includes updating the support and computing the log-likelihood every 50 cycles. Ptychography algorithms update both object and probe, and the time per cycle scales linearly with the number of frames. Note that while powers-of-two sizes are reported here, the FFT libraries used allow sizes with prime number decomposition factors up to 7 (CUDA) or 13 (OpenCL). The CDI speeds are close to optimal, e.g. achieving an average memory throughput of 677 GB/s during a RAAR cycle, but some improvements can still be added for Ptychography algorithms which are more complex, notably for some parts (e.g. position updates, not shown here) which do not achieve high memory throughput. Note finally that for relatively small datasets (less than a minute data processing), the initialisation of scripts (kernel compilations, random number generation on large arrays, ...) and input/output time can require relatively large amounts of time ($>$10 seconds) which affect the overall performance - this can however be mitigated by chaining the analysis of multiple datasets with the same configuration, thus avoiding unnecessary initialisation.}

\end{table}

Examples of speed achieved for various CDI configurations are given in table \ref{table_speed}, and includes the average timing for cycles, the time for updating the support and reporting the log-likelihood every 50 cycles.

The most important features of the CDI module are:

\begin{itemize}
    \item main reconstruction algorithms include: Error Reduction \cite{gerchberg_practical_1972}, Hybrid Input-Output \cite{fienup_phase_1982}, Relaxed Averaged Alternating Reflections \cite{luke_relaxed_2005}, Charge Flipping \cite{wu_reconstruction_2005}, Maximum-Likelihood conjugate gradient \cite{thibault_maximum-likelihood_2012}, General Proximal Smoothing \cite{pham_generalized_2019}
    \item support update function based on a threshold level (relative to the maximum or the support average or root-mean-square), optionally followed by multiple steps of shrinking and expanding by a few pixels \cite{marchesini_x-ray_2003}
    \item taking into account partial coherence using a point-spread function convolution kernel \cite{clark_high-resolution_2012}
    \item initial support determination using either a fixed geometrical form or auto-correlation \cite{marchesini_x-ray_2003}
    \item detwinning algorithms \cite{guizar-sicairos_understanding_2012}
    \item read and write data or reconstructed object using the Coherent X-ray Imaging (CXI) format \cite{maia_coherent_2012} which is based on hdf5, also using NeXus formatting \cite{klosowski_nexus:_1997}. This in turn allows the automatic display of relevant data when opening the files with the silx toolkit \cite{silx_silx-_2020, vincent_silx-_2020}, as shown in Fig. \ref{fig-result-script-ptycho-cdi}b)
    \item functions to simulate data, both for testing and educational purposes.
    \item using the 'free-log-likelihood' figure of merit which provides an unbiased metric for CDI \cite{favre-nicolin_free_2020} and allows the automatic selection of the best object estimate without \textit{a priori} knowledge of its support \footnote{the complete analysis code used for the figures of that article is available (along with datasets) from \cite{favre-nicolin_free_2019}, and can be used as examples of CDI analysis with PyNX.}
    \item automated testing functions to check the code (including consistency between OpenCL and CUDA calculations)
\end{itemize}


\section{\label{sec_ptycho}Ptychography}

Ptychography was first developed for electron microscopy \cite{hoppe_principles_1970,hoppe_trace_1982,rodenburg_phase_2004} and then exploited for coherent X-ray microscopy \cite{rodenburg_hard-x-ray_2007, thibault_high-resolution_2008}. The technique relies on the coherent scattering of an extended object with a shifting illumination. Exploitation of the redundancy of the overlapped illuminated areas yields robust reconstructions with a variety of algorithms \cite{thibault_probe_2009,maiden_improved_2009,thibault_maximum-likelihood_2012, stefanov_efficient_2014, marchesini_alternating_2016, odstrcil_iterative_2018}, for which several software packages are available \cite{nashed_parallel_2014, marchesini_sharp-_2016, enders_computational_2016, dong_high-performance_2018}.

The implementation of the \verb ptycho  module in PyNX follows the same principle as for the CDI one, with a separation between data and the mathematical operators, with three types of objects:
\begin{enumerate}
    \item a \verb PtychoData  object includes all experimental parameters: observed intensity, probe translation positions, detector distance, X-ray wavelength, bad pixel mask, near or far field flag
    \item a \verb Ptycho  object includes a \verb PtychoData  object, the current object and probe estimates (with optionally several modes \cite{thibault_reconstructing_2013}), an incoherent background, and an array $\Psi_j(\mathbf{r})$ which is the view of the multiplication of the object and probe at different positions, and can be propagated from sample to detector space.
    \item a family of Ptychography \textit{operators}, each one allowing to alter a \verb Ptycho  object using the same left-multiplication approach. An operator may alter either object, probe, $\Psi$ arrays, or perform another task (display, export..)
\end{enumerate}

Note that in practice the $\Psi$ array is set to a fixed number of N frames, typically a stack between 16 to 128 (it does not need to be a power of two) but larger values can be used, which are simultaneously computed to achieve higher performance through parallelism, while avoiding excessive memory usage. Operators will then loop over the stack of frames to take into account the entire dataset.

The operators can be used in exactly the same way as for CDI, e.g. doing 100 cycles of difference map and then maximum likelihood on a \verb Ptycho  object \verb p  would be written:
\begin{verbatim}
  p =  ML(update_obj=True,update_probe=True)**100 * \
       DM(update_obj=True,update_probe=True)**100 * p
\end{verbatim}

Most of the high-level operators include a number of options which tailor the calculations accordingly, e.g. for the \verb ML  operator, the full list with default values is:
\begin{verbatim}
  ML(update_object=True, update_probe=False,
     update_background=False, floating_intensity=False,
     reg_fac_obj=0, reg_fac_probe=0,
     calc_llk=False, show_obj_probe=False, fig_num=-1,
     update_pos=False, update_pos_mult=1,
     update_pos_max_shift=2, update_pos_history=False)
\end{verbatim}
These options control not only what is optimised but also log-likelihood printing and graphical display.

\begin{figure}
\label{fig-mpiptycho-modulator}
\includegraphics[width=\textwidth]{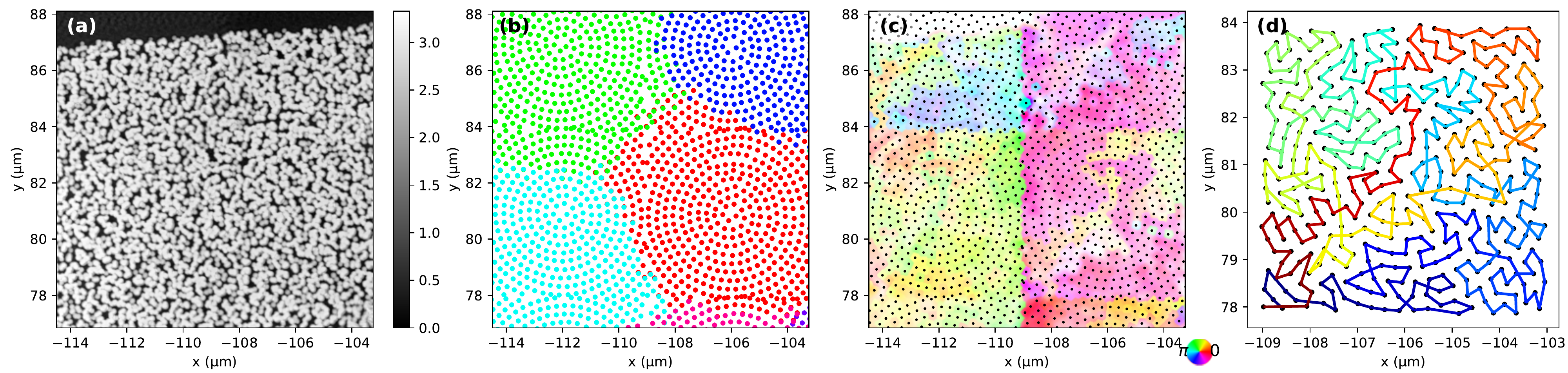}
\caption{Example Ptychography distributed analysis MPI. The object array is about $3000\times3000$ pixels, $30\times30\mu m^2$ with a pixel size of 10 nm, and was reconstructed from 6560 scan positions, composed of 16 scans of 410 points (only 6 are partially visible in the shown area). (a) phase of a $11\times11\mu m^2$ area of the object. (b) distribution of the scanning positions between the 12 GPU - note that some points are shared between neighbouring sets of points, for synchronisation. (c) heat map of the position optimisation, where the colour indicates the direction, and the saturation indicates the amplitude of the displacement relative to the maximum (153 nm). (d) the order with which the scanning positions are visited to minimise motor displacements, following the line from red-$>$yellow-$>$green-$>$blue. This scanning order is visible in (c), where the line corresponding to the red part of the scan in (d) is lighter than the surrounding points, which is particularly visible in the top right and center right scans. Note that after stitching, the borders between the different distributed parts in (b) are not visible in either (a) or (d), which indicates that both phases and positions have been correctly synchronised. Also note at the top left of in (a) that there is no contrast in the object, which is why the position updates (c) have been inhibited in this area due to the lack of contrast. A wider view of (a), (b) and (c) is available as supplementary figures \ref{fig:suppl5} and \ref{fig:suppl6} }
\end{figure}

The most important features of the \texttt{ptycho} module are:

\begin{itemize}
    \item main reconstruction algorithms include: Difference Map (DM) \cite{elser_searching_2007}, Maximum Likelihood (ML) conjugate gradient (based on Poisson noise) \cite{thibault_maximum-likelihood_2012}, Alternating Projections (AP) \cite{marchesini_augmented_2013}
    \item near and far-field geometries
    \item position optimisation (during AP and ML) following the method presented by Odstr\v cil \textit{et al} \cite{odstrcil_iterative_2018}. Note that as this method relies on comparing the shift of the back-propagated exit wave to the object gradient, a filter has been added to avoid shifting the positions when the norm of the object gradient is too small. See an example application if Figure \ref{fig-mpiptycho-modulator}
    \item incoherent background optimisation (AP) \cite{marchesini_augmented_2013}
    \item floating intensities (currently only in OpenCL for AP and ML algorithms) \cite{enders_computational_2016}
    \item multiple incoherent modes for the probe \cite{thibault_reconstructing_2013}
    \item smoothing parameters for object and probe update (via regularisation for ML, and inertia with a convolution kernel for AP and DM) \cite{enders_computational_2016}
    \item functions to simulate data, both for testing and educational purposes
    \item reporting of log-likelihood based on Poisson, Gaussian and Euclidian noise models \cite{thibault_maximum-likelihood_2012}
    \item read and write data or reconstructed object and probe using the hdf5-based Coherent X-ray Imaging format \cite{maia_coherent_2012}
    \item automated testing functions to check the code (including consistency between OpenCL and CUDA calculations)
\end{itemize}

Up to early 2020, ptychography datasets collected at ESRF did not require large amounts of GPU memory, and could be analysed with only 16 or 32 GB. However that is quickly changing with new synchrotron sources which can produce a much higher coherent flux \cite{johansson_max_2016, raimondi_esrf-ebs:_2016, schroer_hard_2016}. It is therefore useful to exploit multiple GPU or computing nodes \cite{nashed_parallel_2014, enders_computational_2016, dong_high-performance_2018} to handle larger ptychography datasets.

PyNX now includes the ability to analyse datasets with multiple GPU and/or multiple computing nodes, using the Message Passing Interface through the \textit{mpi4py} python module \cite{dalcin_mpi_2005, dalcin_parallel_2011}. For this, a new \verb#PtychoSplit#  class has been derived from the \verb#Ptycho#  one, which manages the coordination between all processes. This follows the \textit{asynchronous} approach presented by Nashed \textit{et al}\cite{nashed_parallel_2014}, minimising latency due to the synchronisation (as object and probe arrays need to be copied from the GPU to the host memory, and then between compute nodes).

The process follows the different steps:
\begin{enumerate}
    \item the script (see section \ref{sec_cli}) loads the scan positions, and distributes them among the different process. This is done by using the k-means algorithm from the \verb#scikit-learn# python module \cite{pedregosa_scikit-learn_2011}, after which the distribution of scanning positions is adjusted between neighbour sets to reach an homogeneous number of positions per process, and finally every set has to share at least 20 scanning positions common to neighbouring sets, for later synchronisation.
    \item each process loads the images
    \item object and probe are initialised in the master process and each part is distributed among the different process
    \item the analysis algorithm is performed independently among the different process, as for a normal ptychography analysis
    \item the different parts of the object are stitched together (this can also be done during the algorithm e.g. if the user requests a graphical update of the object).
\end{enumerate}

The stitching step has to take into account the different ways the object and probe can differ between independent processes, as only their multiplication can be quantified, unless an image without the object has been included in the dataset. Thus object and probe can have different relative scale factors, a different phase shift, and also a different phase ramp (linearly varying over the 2D array dimensions, in opposite ways for the object and probe). Finally, if the scan positions are updated, the average shifts can differ between the parallel processes.

The relative phase ramp of the object and probe is first removed in each process independently using the \verb#ZeroPhaseRamp# operator, which computes the center of mass of the square norm of the Fourier transform of the probe, and then corrects both probe and object for the phase ramp corresponding to the sub-pixel shift relative to the center of the array in Fourier space.\footnote{This is also applied for non-distributed ptychography, so that independent optimisations of the same dataset will yield the same phase ramp for the object.}$
^,$\footnote{Note that this approach can also be used to remove the object phase ramp for far field ptychography: by computing the sum of all calculated diffraction frames and measuring the sub-pixel shift of the center of mass of this intensity, the average phase ramp in the object can be determined. However, as the object can have a varying thickness or composition, this is not an absolute method to derive the correct phase, which can only be obtained if a reference direct beam is used, as shown by Diaz \textit{et a}\cite{diaz_quantitative_2012}.}

The stitching is then done in the following way:
\begin{enumerate}
    \item the probes are aligned (up to a pixel precision)
    \item the scan positions (if they have been updated) are merged and the relative shifts minimised (using the shared positions between neighbouring sets of points)
    \item the phase differences between different  parts are minimised, by using small areas of the object around the shared scanning positions.
    \item the object is finally stitched by a linear combination of the different object parts, weighted by the object illumination for each part.
\end{enumerate}

\begin{figure}
\label{fig-mpiptycho-performance}
\begin{center}
\includegraphics[width=0.7\textwidth]{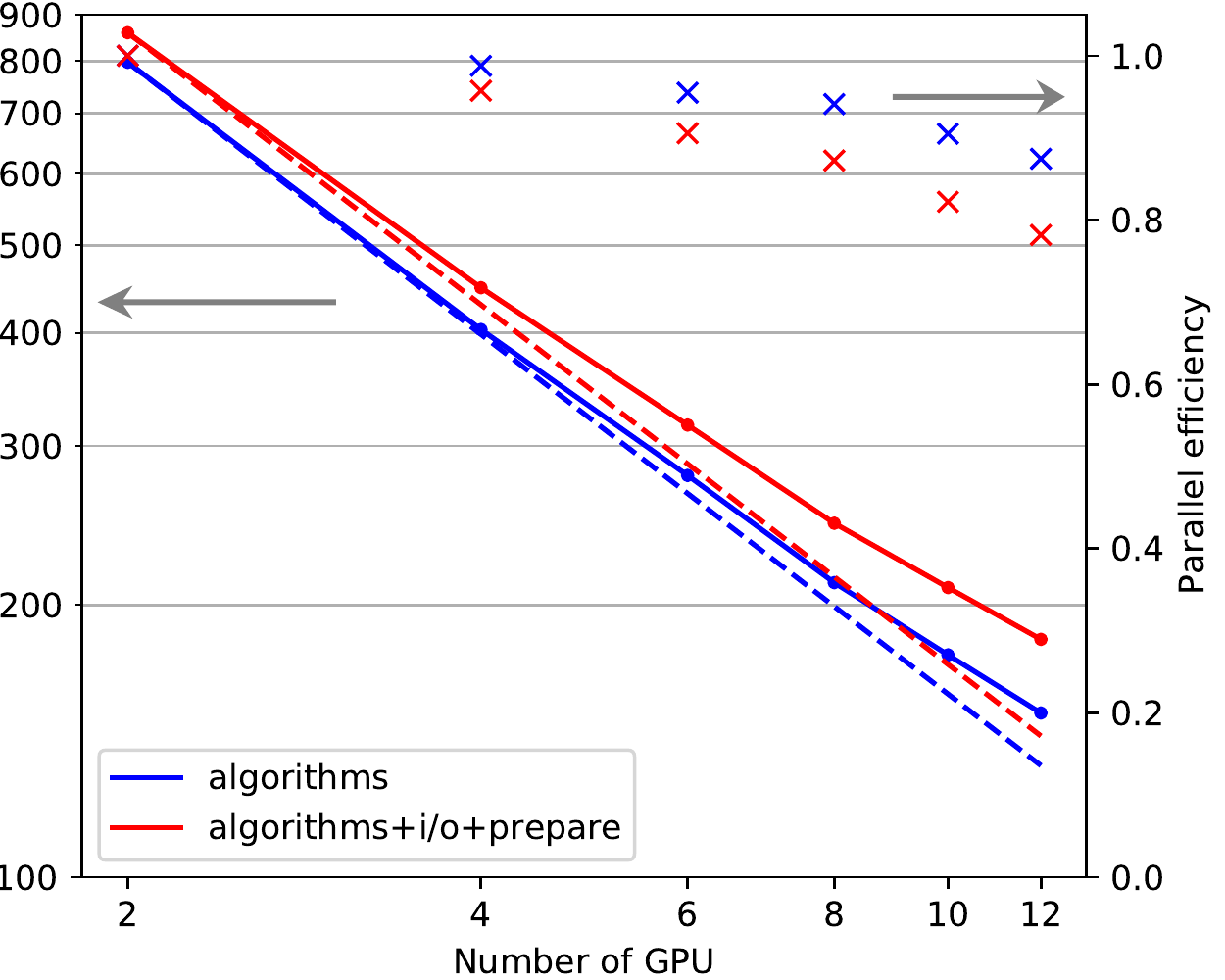}
\end{center}
\caption{Performance of the MPI-distributed ptychography reconstruction presented in Fig. \ref{fig-mpiptycho-modulator} with a dataset comprising 6560 frames with 400$\times$400 pixels, using 1 to 6 compute nodes with two Nvidia v100 GPU each. The blue and red lines indicate the compute time in seconds, either only for the algorithmic parts (including the final object stitching), or also including the input/output and preparation. The relative efficiency of the parallelism is indicated by the crosses, and remains above 87\% (78\% with i/o) for 12 GPUs, even though each GPU is only handling a relatively small dataset with about 570 positions.}
\end{figure}

Note that in principle, this \textit{asynchronous} approach with stitching at the end of the process also allows for a \textit{sequential} analysis of the different parts of the object, which could only be stitched together after all parts have been processed. This would be useful when an insufficiently large enough GPU cluster is available, and would allow to handle very large datasets, as long as the final object does not exceed available memory.

An example result of distributed optimisation is shown in Figure \ref{fig-mpiptycho-modulator}, for a modulator object reconstructed from 6560 scan positions, with diffraction images cropped to $400\times400$ pixels. The analysis was done using 3 probe modes, first with 200 DM cycles, then 100 AP cycles, then 1000 AP cycles with positions updates, then 100 ML cycles.

This analysis was performed on the ESRF GPU cluster using from 1 to 6 compute nodes, each with two Nvidia V100 GPUs with 32 GB of memory. The performance of the parallel calculation is shown in Figure \ref{fig-mpiptycho-performance}, with the overall time (including i/o and object/probe initialisation) going from 860 to 183 seconds using two or twelve GPU. Note that when using 6 nodes, the number of images per GPU is only about 570, which is fairly low and contributes to decreasing the overall parallel efficiency at 78\%.

Other large-scale tests are reported in Table \ref{table_speed}, with 70k $512\times512$ frames and 250k $256\times256$ frames, spread over 12 GPU.


\section{\label{sec_cli}Command-line scripts}

In addition to the Python programming interface, both CDI and Ptychography datasets can be analysed using command-line scripts, either at the beamline during an experiment, or afterwards. These scripts accept parameters which give access to a range of options, as well as specifying the algorithm in a simple mathematical-like string, which can be easily interpreted using the operator approach.
\subsection{CDI}
For example, the analysis of a CDI dataset stored in a CXI file, can be simply written:
\begin{verbatim}
  pynx-id01cdi data=pt.cxi support_threshold=0.2
\end{verbatim}
which would perform a simple analysis based on default parameters (600 cycles of HIO followed by 200 cycles of ER, including a support update and printing the log-likelihood every 50 cycles).

Alternatively, the exact algorithmic chain can be specified using the \verb algorithm  keyword: 
\begin{verbatim}
  pynx-id01cdi data=pt.cxi algorithm=ER**200*HIO**600
               support_threshold=0.2 liveplot
\end{verbatim}

\begin{table}
\begin{tabular}{|c|m{6cm}|m{8cm}|}      
\hline
 Technique    & algorithm chain & details     \\
\hline
CDI & \texttt{ER**50,(Sup*ER**5*HIO**50)**10}& 50 cycles of HIO followed by 5 cycles or ER and support update, repeated 10 times, then 50 cycles of ER\\ \hline
CDI & \texttt{ER**50,(Sup*PSF*HIO**50)**4, positivity=0,(Sup*HIO**50)**8} & 50 cycles of HIO followed by support update, repeated 8 times, then deactivation of a positivity constraint,then 50 cycles of HIO with the Point-Spread-Function partial coherence kernel update and support update, repeated 4 times, then  50 cycles of ER\\
\hline
Ptycho & \texttt{ML**40,DM**100,probe=1}& Activate probe optimisation, then 100 DM and 40 ML cycles, followed by an analysis of the probe (determination of focal point, widths, and mode decomposition)\\ \hline
Ptycho & \texttt{ML**100,DM**200,nbprobe=3,ML**40, DM**100,probe=1,DM**20} & First 20 DM cycles with object update only, then 100 DM and 40ML whilst also updating the probe, then use 3 probe modes and do 200 DM followed by 100 ML cycles\\  \hline
Ptycho  & \texttt{(ML**10*AP**20)**3,position=1, AP**50*DM**50,nbprobe=2,probe=1} & First 50 DM cycles with object and probe update and 2 probe modes, followed by 50 AP and 50 ML, then activate probe position optimisation, then perform 20 AP and 10 ML cycles 3 times\\
\hline

\end{tabular}

\caption{\label{table_algorithm_chain}Example of algorithm chains which can be used for CDI or Ptychography analysis using the command-line scripts. Each chain will be executed from right to left, as when applying an operator to a mathematical array in an equation. Each step can either be a modification of a default parameter (e.g. the \texttt{positivity} for CDI, the number of probe modes (\texttt{nbprobe} ) for ptychography, or a more specific task (\texttt{analysis},)... Alternatively a chain of algorithm operators can be given (\texttt{ER}, \texttt{DM}, etc..). After each step which is comma-separated, the result can be saved to a CXI file. This approach allows for a great flexibility of the algorithm, without any compromise on the performance since all optimisation steps are queued asynchronously on the GPU.}
\end{table}

This approach allows the full customisation of the sequence of algorithms used, mixing all standard algorithms (ER, HIO, RAAR,...) along with other algorithms (support update, partial coherence) or parameters (positivity constraint..). More examples of how the \texttt{algorithm=} keyword can be used to customise the actual algorithm chain is given in table \ref{table_algorithm_chain}. 

As indicated by the name -\texttt{pynx-id01cdi}- this script is tuned for the ESRF id01 beamline \cite{leake_nanodiffraction_2019}, with the default parameters optimised for Bragg CDI. Another script -\texttt{pynx-id10cdi}- is also available, with different parameters, notably disabling the use of the auto-correlation to determine the initial support as on the ESRF id10 beamline, a central stop is used, making that method ineffective. Other customised scripts (e.g. to handle different types of input files) can easily be added.

These scripts can exploit multiple GPU on one or multiple nodes to either distribute the analysis of multiple scans, or when multiple analysis runs are performed on the same dataset (e.g. to select the best solution based on the free log-likelihood analysis \cite{favre-nicolin_free_2020}) by performing the calculations on any number of parallel processes.

Many other options are available for the command-line scripts, as listed in the online documentation \footnote{http://ftp.esrf.fr/pub/scisoft/PyNX/doc}.

\subsection{Ptychography}

The scripts for ptychography analysis follow the same principles, for example when reading a CXI data file (which includes all observed frames, probe positions, detector distance, mask, wavelength), one can use:
\begin{verbatim}
 pynx-cxipty.py data=data.cxi 
  probe=focus,60e-6x200e-6,0.09
  algorithm=analysis,ML**100,DM**200,nbprobe=3,probe=1 
  saveplot liveplot
\end{verbatim}

This will trigger the data analysis, starting from simulating the initial probe as a rectangular aperture of $60(h)\times 200(v)\ \mu m^2$ focused by a lens with a focal length of 9 cm, then optimising the object and the probe (3 modes) with 200 cycles of DM and 100 cycles of ML, followed by an analysis of the resulting probe. The \verb#saveplot# option allows to save images, including a view of both object and probe as depicted in Fig. \ref{fig-result-script-ptycho-cdi}a), or of the probe analysis results (width, propagation to the focal point, modes), and a map of the scan position shifts like in Fig. \ref{fig-mpiptycho-modulator}c) if these were optimised.

Other scripts are available to handle directly data from different beamlines (id01, id13 and id16A at ESRF, NanoMAX at MaxIV, Nanoscopium and Cristal at Soleil), or from different software such as PtyPy \cite{enders_computational_2016}. These scripts all use the same base code, the main change being the functions to load data from the various input files and formats, so that it can easily be extended to other input formats.


\section{Conclusion}

To conclude, the PyNX toolkit provides a wide range of modules for the simulation and analysis of coherent imaging data, which transparently exploits accelerated computing on GPUs using either the OpenCL or CUDA language. The programming approach, which mimics mathematical operators, affords a great flexibility in the choice of algorithm chains to be used for data analysis.

Command-line scripts are also available to handle CDI and ptychography datasets without any programming knowledge, and new ones can easily be added. 

PyNX is open-source (CeCILL-B license\footnote{https://cecill.info/} similar to the BSD one) and freely available, distributed by ESRF from http://ftp.esrf.fr/pub/scisoft/PyNX/. This includes installation scripts (available for Linux and macOS) to create python virtual environments with all necessary dependencies. The online documentation includes a number of examples as jupyter notebooks. The git repository is also accessible on the ESRF gitlab server (https://gitlab.esrf.fr) on demand.

\section{\label{appendix_cdi_operators}CDI operators and example code}

The list of the main CDI operators is given below. All operations (with the exception of graphical display) are executed transparently on the GPU, and are implemented both for the OpenCL and CUDA languages.

\begin{tabular}{|C{0.3\textwidth}|C{0.7\textwidth}|}      
\hline
 Name    & CDI Operator     \\
\hline
 \verb#AutoCorrelationSupport# & Initialise the object support from the intensity auto-correlation \cite{marchesini_x-ray_2003} \\ \hline
 \verb#FT#       & Fourier Transform              \\ \hline
 \verb#IFT#      & Inverse Fourier Transform              \\ \hline
 \verb#ApplyAmplitude#       & In Fourier space, replace the modulus by the observed amplitude              \\ \hline
 \verb#FourierApplyAmplitude#       & \verb@IFT() * ApplyAmplitude() * FT()@ \\ \hline
 \verb#ERproj#       & Set the object to zero outside the support (support projection) \\ \hline
 \verb#ER#       & \verb@ERproj() * FourierApplyAmplitude()@ \\ \hline
 \verb#EstimatePSF#       & Update the point-spread-function kernel to take into account partial coherence \cite{clark_high-resolution_2012} \\ \hline
 \verb#HIO#       & Hybrid Input-Output (also uses \verb#FourierApplyAmplitude#) \\ \hline
 \verb#LLK#      & Compute the Poisson log-likelihood from the calculated and observed intensities \\ \hline
 \verb#PRTF#      & Compute and plot the Phase Retrieval Transfer Function \cite{chapman_high-resolution_2006, chushkin_three-dimensional_2014} \\ \hline
 \verb#RAAR#       & Relaxed Averaged Alternating Reflections (also uses \verb#FourierApplyAmplitude#) \\ \hline
 \verb#ShowCDI#       & Plot the current estimate of the object (amplitude, phase) with the observed and calculated amplitude\\ \hline
 \verb#SupportUpdate#       & Update the support based on a threshold relative to the maximum amplitude in the object 
 optionally expanding or shrinking the support afterwards \cite{marchesini_x-ray_2003}\\ \hline
\end{tabular}

\section{\label{appendix_ptycho_operators}Ptychography operators}

The list of the main Ptychography operators is given below. All operations are executed transparently on the GPU, and are implemented both for the OpenCL and CUDA languages. Some operators actually apply to a stack of N frames (N=16 to 128), whereas others apply to all frames by looping over all stacks of frames.

\begin{tabular}{|c|m{0.7\textwidth}|}      
\hline
 Name    & Ptychography Operator     \\
\hline
 \verb#AnalyseProbe# & Analyse the probe (modes, determination of focus and width) \\ \hline
 \verb#ApplyAmplitude#       & In Fourier space, replace the modulus by the observed amplitude              \\ \hline
 \verb#AP#       & Alternating projection algorithm \\ \hline
 \verb#DM#       & Difference Map algorithm \\ \hline
 \verb#FT#       & Fourier Transform              \\ \hline
 \verb#IFT#       & Inverse Fourier Transform              \\ \hline
 \verb#LLK#      & Compute the Poisson log-likelihood from the calculated and observed intensities \\ \hline
 \verb#ML#       & Maximum Likelihood (conjugate gradient, Poisson noise) algorithm \\ \hline
 \verb#ObjProbe2Psi# & Multiply object and probe, $\Psi_j(\mathbf{r})= O(\mathbf{r})P(\mathbf{r-r_j})$ \\ \hline 
 \verb#PropagateApplyAmplitude#       & \shortstack{ \texttt{ IFT() * ApplyAmplitude() * FT() } or \\ \footnotesize \texttt{ PropagateNearField() * ApplyAmplitude() * PropagateNearField()} }\\ \hline
 \verb#PropagateNearField# & Near field propagation (forward or backward) \\ \hline
 \verb#Psi2ObjProbe#       & Update the object and/or probe from the back-propagated $\Psi_j$\\ \hline
 \verb#Psi2PosShift#       & Update the shift of illumination positions \cite{odstrcil_iterative_2018} \\ \hline
 \verb#ShowObjProbe#       & Plot the current estimate of the object (amplitude, phase) and probe\\ \hline
\end{tabular}

The authors acknowledge the help of Manfred Burghammer, Julio Cesar da Silva, Virginie Chamard, Peter Cloetens, Jo\"el Eymery, Tilman Gruenewald, Ross Harder, St\'ephane Labat, Kadda Medjoubi, Linus Pithan and Tobias Sch\"ulli for useful discussions and/or test datasets. A number of features included in PyNX are inspired from the open-source package PtyPy \cite{enders_computational_2016}, as well as from the CDI matlab code originally written by Jesse Clark.



\appendix
\section{Appendix}
\subsection{OpenCL vs CUDA FFT performance}

Both OpenCL and CUDA languages rely on the same hardware. Generally speaking, the performance is almost identical for floating point operations, as can be seen when evaluating the scattering calculations (Mandula et al, 2011). However the FFT performance depends on low-level tuning of the underlying libraries, namely the cuFFT and clFFT libraries, which are respectfully optimised for Nvidia and AMD devices.

The performance of both libraries has been evaluated for an Nvidia V100 GPU, for 2D and 3D FFT of all sizes for which the largest prime factor decomposition is at most 7 (note that clFFT allows up to 13, and cuFFT allows values larger than 7 but with a degraded performance). The configuration used for the comparison was: Nvidia driver 435.21, CUDA version 10.1, clFFT v2.12.2, pyopencl 2019.1.2, pycuda 2019.1.2, gpyfft git commit 2c07fa8e7674757.

\begin{figure}[H]
    \centering
    \includegraphics[width=\textwidth]{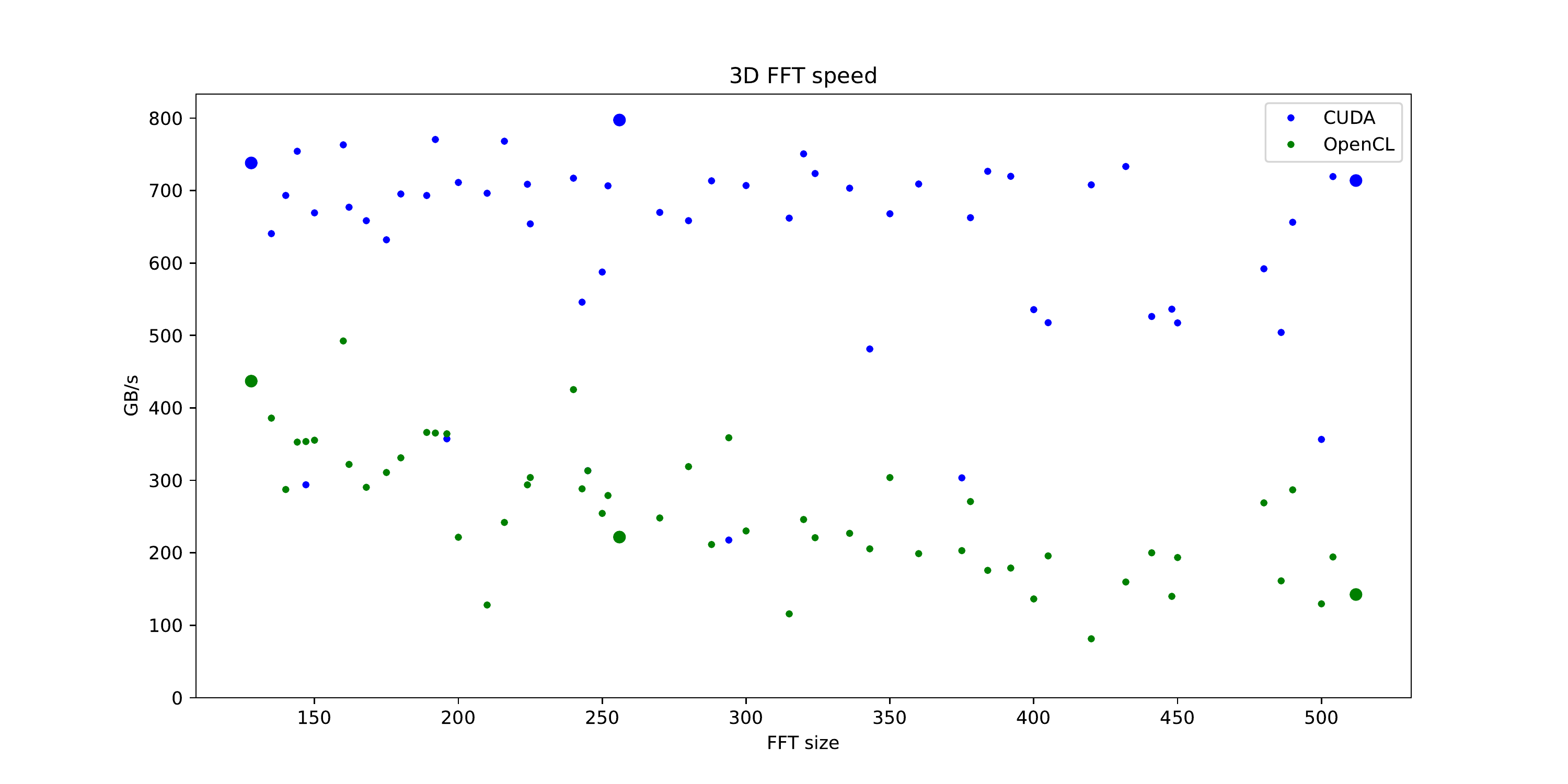}
    \caption{3D FFT performance, measured on a Nvidia V100 GPU, using CUDA and OpenCL, as a function of the FFT size. The obtained speed can be compared to the theoretical memory bandwidth of 900 GB/s. Larger dots are shown for power-of-twos transforms}
    \label{fig:suppl1}
\end{figure}

\begin{figure}[H]
    \centering
    \includegraphics[width=\textwidth]{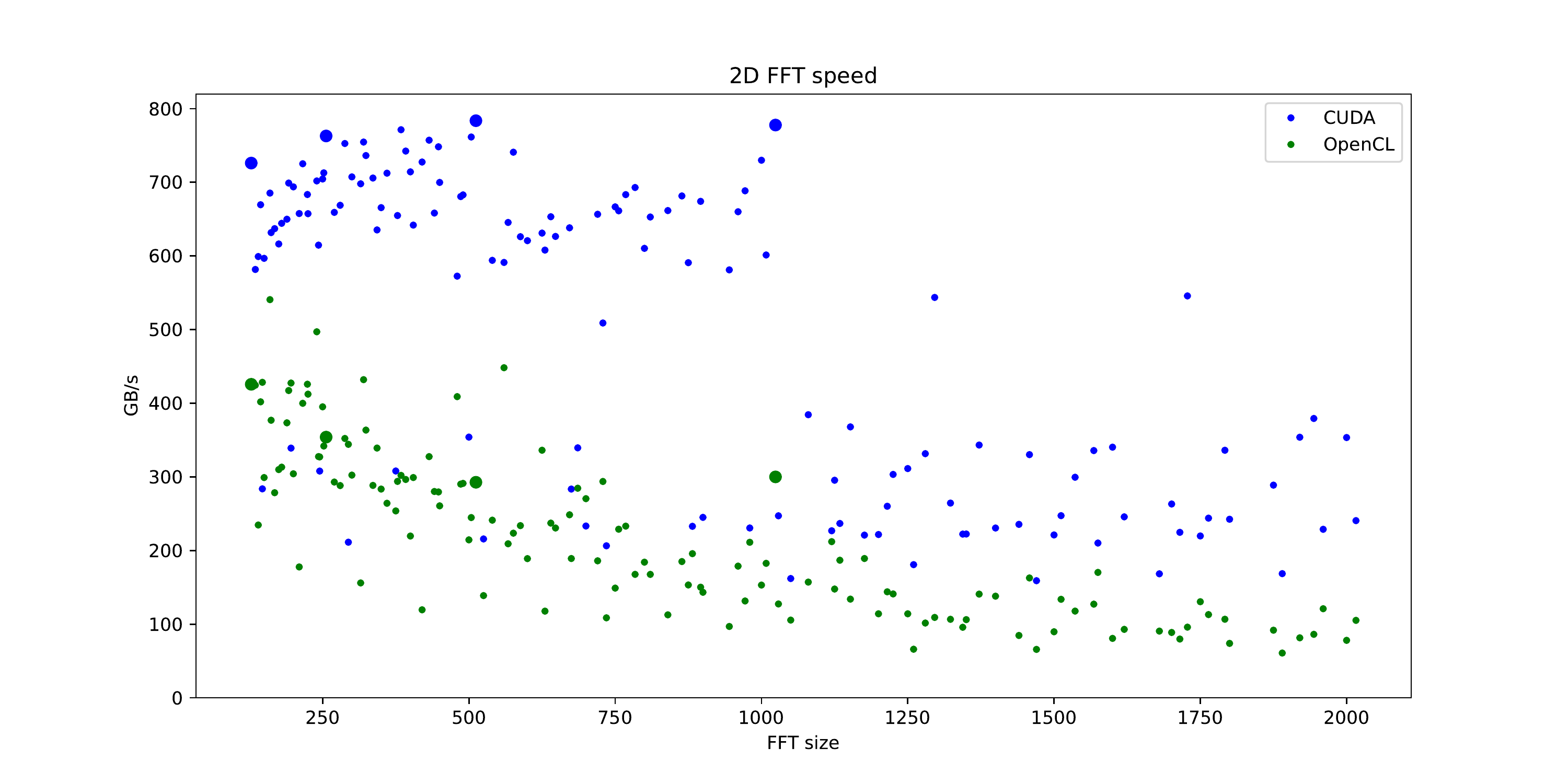}
    \caption{2D FFT performance, measured on a Nvidia V100 GPU, using CUDA and OpenCL, as a function of the FFT size up to N=2000. The obtained speed can be compared to the theoretical memory bandwidth of 900 GB/s. Larger dots are shown for power-of-twos transforms}
    \label{fig:suppl2}
\end{figure}

As performance on a GPU is limited by the memory throughput rather than the floating-point operations, we report in Fig. \ref{fig:suppl1} and \ref{fig:suppl2} the average processing speed in GB/s, taking into account the N read and write operations for the N-dimensional FFT. Each test is done by performing two pairs of backward and forward FT in single precision (32-bit floating point), and the test is repeated four times, the best time being kept for reporting. In the case of clFFT, each possible order for the axes transforms (a N-dimensional FT is a succession of N 1-dimensional FT) for the FT is tested and the best time is used.

Note that these tests do not imply that cuFFT is superior to clFFT \textit{in general}, but rather that it is at least the case \textit{on Nvidia hardware}. This is expected as clFFT is optimised for AMD GPU. One notable difference is the warp-size which is 32 for Nvidia GPU, whereas for AMD the wavefront is 64 – both numbers correspond to the number of low-level compute threads which are executed in parallel on a compute unit – a difference which can explain that clFFT tuning is not optimal on Nvidia hardware.
Also note that ‘in-place’ cuFFT transforms require 2x the amount of memory for the transform, in order to optimise memory transfers (it is faster to copy values from adjacent memory, so the transforms along the different axis of the FFT are better optimised by re-arranging the memory). clFFT does not require this (which may be a reason for a lower performance), and can thus handle larger transforms, which can be useful for large 3D CDI FFT.
All the tests can be reproduced using the function:
 \verb#pynx.test.speed.plot_fft_speed()#

\subsection{Larger version of Figures 5a and b}
\begin{figure}[H]
    \centering
    \includegraphics[width=0.9\textwidth]{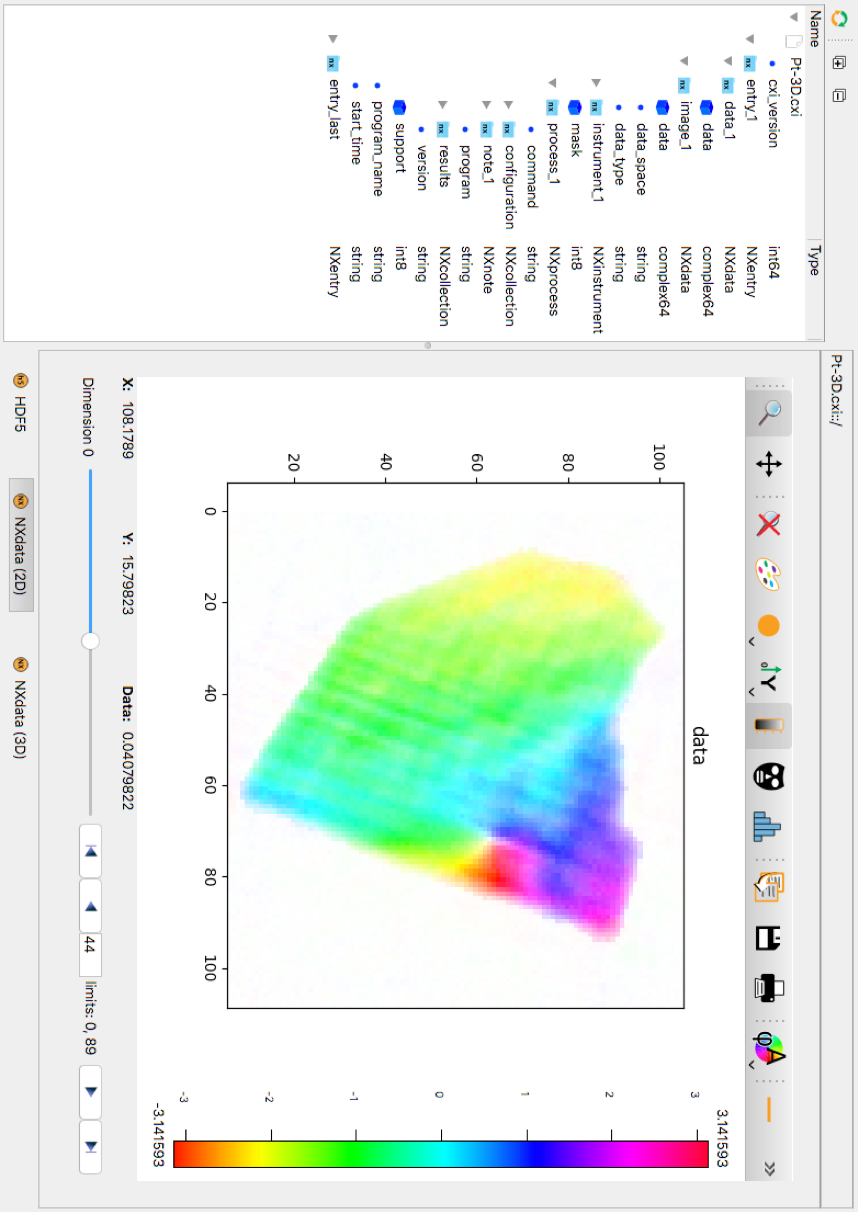}
    \caption{larger version of article Figure \ref{fig-result-script-ptycho-cdi}b. Display of the CXI file, using the silx toolkit viewer, obtained as a result from a CDI analysis. As the CXI file is NeXus-formatted, the view automatically opens the relevant object. The HSV colour map gives information both about the amplitude and the phase of the object. This example is the result of Bragg CDI on a Pt nano-crystal with a dislocation. As can be seen in the left of the image, different fields in the CXI/hdf5 file include information about the object as well as the process and parameters used for the analysis.}
    \label{fig:suppl3}
\end{figure}

\begin{figure}[H]
    \centering
    \includegraphics[width=0.9\textwidth]{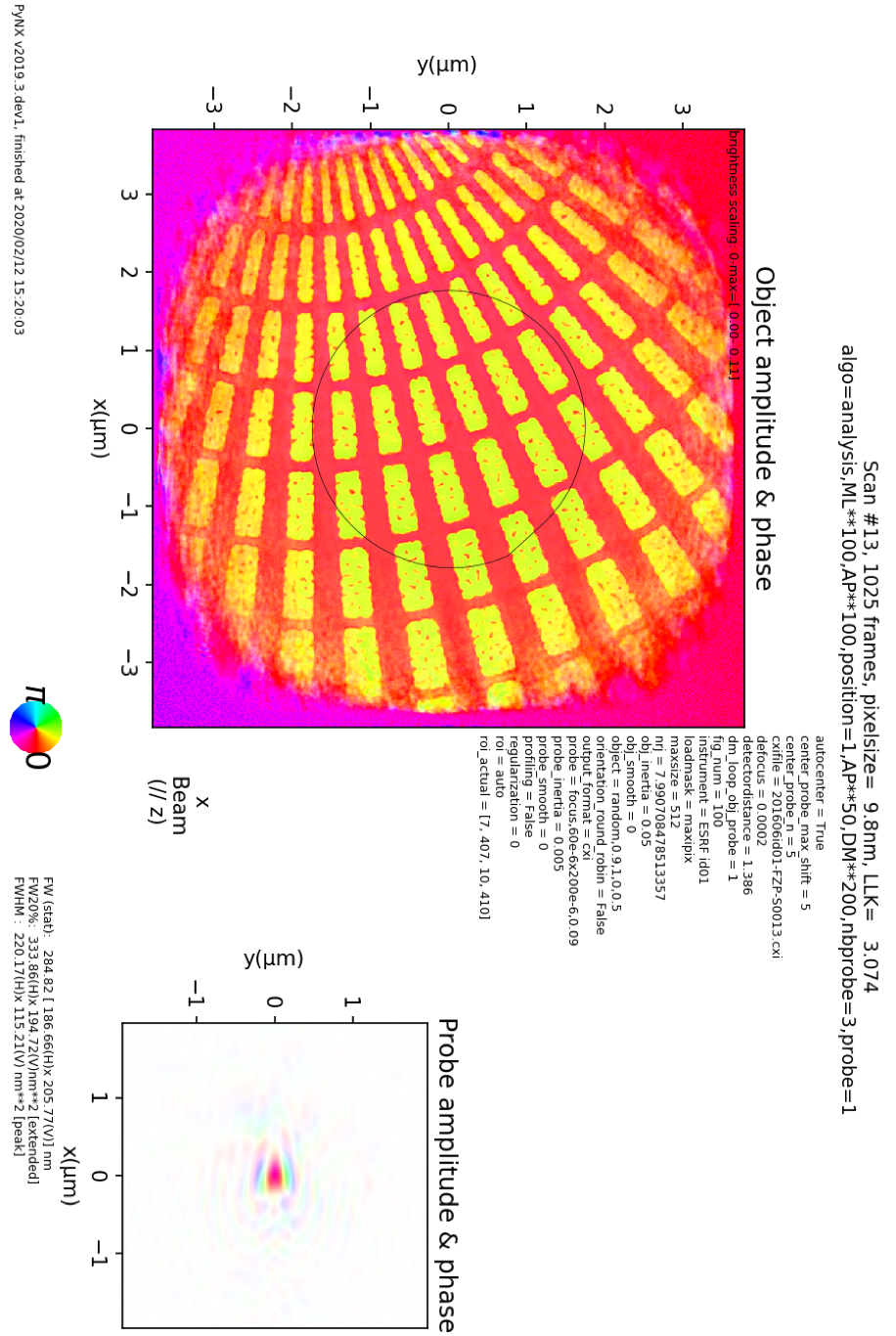}
    \caption{larger version of Fig.\ref{fig-result-script-ptycho-cdi}a - example output plot of a ptychography experiment, showing both the object and the probe in RGBA, as well as all parameters used for the analysis}
    \label{fig:suppl4}
\end{figure}

\subsection{Larger view of the MPI-ptychography reconstruction of the modulator}

\begin{figure}[H]
    \centering
    \includegraphics[width=\textwidth]{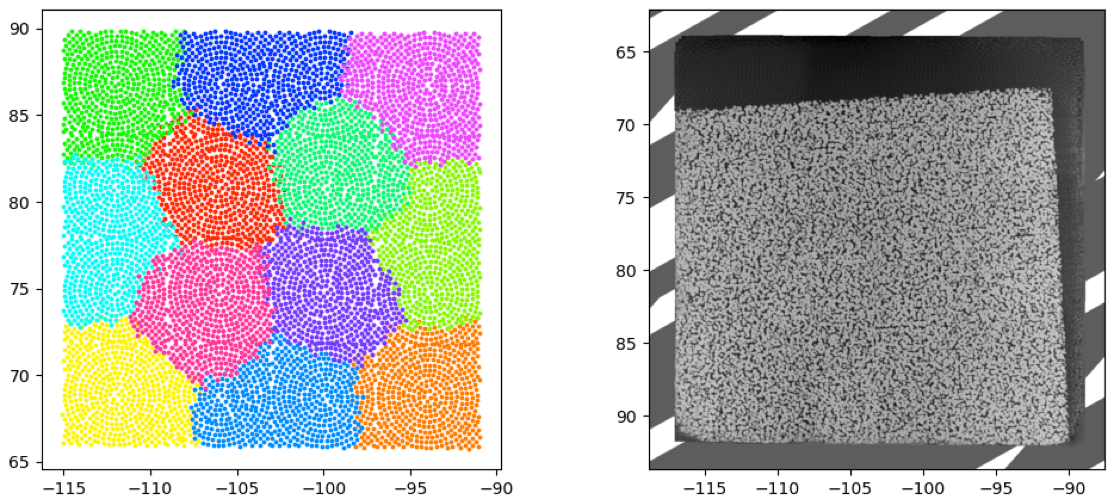}
    \caption{overview of the reconstructed modulator, shown in the article Fig. \ref{fig-mpiptycho-modulator}. Left: Division of the scanning positions into 12 domains with ~575 points each, including ~30 shared with the neighbours. Right: complete view of the reconstructed object phase}
    \label{fig:suppl5}
\end{figure}

\begin{figure}[H]
    \centering
    \includegraphics[width=\textwidth]{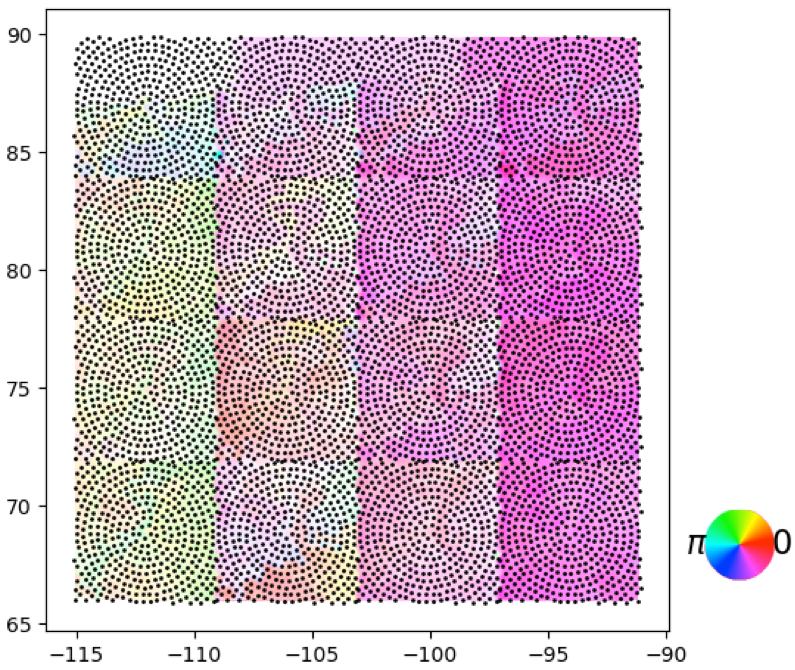}
    \caption{complete heat map of the optimised position shifts (see article Fig. \ref{fig-mpiptycho-modulator}c for explanations), which shows that the overall trend is a drift vs time for the 16 successive scans which compose the entire dataset. In this representation, the maximum displacement is 313nm, and the reference region at the top left.}
    \label{fig:suppl6}
\end{figure}

\end{document}